\documentclass[aps, prd, amsmath, floats, floatfix, twocolumn, nofootinbib, 
]{revtex4}
\usepackage{graphicx}
\usepackage{latexsym}
\usepackage{color}
\usepackage{ dsfont }
\usepackage{amsfonts,amsbsy}
\usepackage[utf8]{inputenc}
\usepackage[mathscr]{euscript}
\usepackage{ amssymb }
\usepackage{amsthm}
\usepackage{amsmath}
\usepackage[T1]{fontenc}
\usepackage[utf8]{inputenc}
\usepackage{mathtools}
\usepackage{physics}
\usepackage[utf8]{inputenc}
\usepackage[english]{babel}

\newcommand{\be}{\begin{eqnarray}}
\newcommand{\ee}{\end{eqnarray}}

\theoremstyle{definition}

\bibliography{mybib}

\begin{document}

\title{A Firepoint at the Black Hole Singularity} 

\author{Ali Akil}
\email{akila@mail.sustc.edu.cn}

\author{Oscar  Dahlsten}
\email{dahlsten@sustc.edu.cn}

\author{Leonardo Modesto}
\email{lmodesto@sustc.edu.cn}

\affiliation{$^{*\, \dagger \, \ddagger}$Department of Physics, Southern University of Science and Technology (SUSTech), Shenzhen 518055, China}

\affiliation{$^\dagger$Wolfson College, University of Oxford, Linton Rd, Oxford OX2 6UD, United Kingdom}

\affiliation{$^\dagger$London Institute for Mathematical Sciences,
35a South Street, Mayfair, London, W1K 2XF, United Kingdom}

\begin{abstract}
In his original derivation, Hawking showed that a Schwarzschild black hole is unstable at quantum level and it evolves to a final thermal mixed state violating unitarity. 
  There are some attempts to solve this information paradox based on a high energy surface located at the black hole event horizon: the firewall. In the wake of these proposals, we here propose the singularity itself as a ``firepoint'' capable to break the entanglement between the ``int'' and the ``out'' states created through the Hawking process. In this paper the singularity takes active part in the information paradox in two similar ways.  In the first way, the singularity, coming in contact with the the ``int'' state, produces a pure state outside the horizon, but it violates causality allowing people inside the black hole to send signals to the outside. The second way consists of a map that breaks the entanglement between the interior and exterior of the black hole still using the singularity yet without violating causality. The monogamy theorem is not violated whether in the first where the ``out'' radiation state is rendered pure directly or in the second way where the Page idea is made possible again to solve the information loss problem.

\end{abstract}

\maketitle

\section{Introduction}
At classical level black holes are stable in a large class of gravitational theories with in primis Einstein's gravity.
However, at quantum level they turn out to be unstable due to the Hawking evaporation process \cite{Hawking}. 
As long as the black hole does not reach its last evaporation stage, it can be proved that emitted radiation at $I^+$ (future infinity) is correlated with radiation carrying negative energy and falling inside the black hole (see for example \cite{Fabbri} for an extended and detailed discussion.) This in general is not problematic, but a paradox emerges at the end of the evaporation process when there is no more black hole, no singularity, but a thermal state of particles at the Hawking temperature $T_{\rm H} =\frac{ 1}{8 \pi M}$ (where $M$ is the mass of the black hole and we are using natural units.) Therefore, in the whole process of collapse and evaporation, a pure state (in an ideal simplified circumstance) has evolved into a mixed one, violating unitarity. 
To solve this issue, D. Page argued that the radiation emitted later must be maximally correlated with radiation emitter earlier in order to end up with a pure state after the full black hole evaporation \cite{Page}. This approach is very conservative from the general relativity point of view because it does not consider any modification of the black hole geometry due to possible quantum gravity corrections that most likely will change the spacetime metric at least at the Planck scale. Indeed, the whole process turns out to be unitary, only looking at the exterior of the black hole. 
However, many years later it was proved that the Page argument could be incorrect because it seems to violate the monogamy of entanglement entropy \cite{Mathur}. Monogamy states that a system A which is maximally entangled with a system B cannot be simultaneously entangled with another system C \cite{coffman}.
The proof of this principle relies on a property of entanglement entropy called strong sub-additivity.
Therefore, if we want to avoid inconsistencies it is argued that we are forced to introduce a ``firewall'', which somehow destroys the entanglement of the particle pair at the event horizon, 
leaving the late and early radiation entangled in order to save the purity of the final state \cite{Braunstein, AMPS}.
This last idea - although sometimes presented as the most conservative
solution to the information paradox [6] - actually breaks
the fundamental equivalence principle on which General Relativity
is based. Moreover it necessitates introducing an extra mechanism 
implementing the firewall, which is a priori not there.

In fact, there are reasonable models in which there is no drama at the horizon, 
and an infalling observer would effectively see information `erased' 
only at the singularity \cite{Thorac1, Thorac2}. Attempts have been made to 
augment and add mechanisms to such arguments such that information is not erased according to 
outside observers \cite{Complementarity, Thorac1, Thorac2, Iran}. In these attempts it is argued that this non-unitary evolution at the singularity has no observable effect.

   Now, we would like to recall the statement of the paradox as given in AMPS \cite{AMPS}: 

{\em [...] in brief, the purity of the Hawking radiation implies that the late radiation is fully entangled with the early radiation, and the absence of drama for the infalling observer implies that it is fully entangled with the modes behind the horizon.This is tantamount to cloning}.

In this paper we provide a story of what possibly happens if one decides to provide a conservative solution of the issue pointed out in \cite{Mathur, AMPS} without introducing any exotic large scale modification of gravity and/or the spacetime causal structure, but takes the singularity seriously.
Indeed, we show that the black hole singularity predicted by General Relativity can play the role of a firewall, or actually a ``firepoint''. For an infalling observer, the latter naturally resets the state falling inside the black hole leaving the ingoing and the outgoing radiation pairs disentangled.
Therefore, we basically admit one kind of firewall, but we assume that there is no need for an extra entity introduced by hand. Indeed, general relativity already provides for us a firewall as an exact solution of Einstein's gravity, namely {\em the zero radius singularity}.

\section{The fate of a quantum state falling into the black hole}
In this section we argue that the quantum entanglement between the particle pair created at the event horizon is actually broken when the infalling particle reaches the singularity.
Let us assume that a pair of particles is created at the horizon, one goes inside the black hole while the other flees away towards the future infinity.
For the sake of simplicity we here consider the Schwarzschild black hole that shows up an essential singularity in $r=0$. Notice that when an object crosses the black hole horizon, it ends up very quickly at the spacetime singularity (e.g. $10^{-6} s$ for a solar mass black hole). The singularity is -as is well known- a very special point in spacetime having two main properties: (i) the spacetime is not extendible beyond $r=0$ which is to say any particle is forced to end up there, (ii) any local curvature invariant is divergent in $r=0$ \cite{Landau}.
Therefore, whether a particle enters a black hole its position is shortly determined with extremely high precision. Since our intention here is to take the singularity seriously, we could say that the position of the particle is known with infinite accuracy. 
The ``Schwarzschild time $t_{\rm s}$'', is defined as the time needed for a particle to reach the singularity starting from a region nearby the event horizon and it equals (see the appendix)
\be \label{timesing}
t_{\rm s} = 2 r_{\rm s}/3 \sim r_{\rm s},
\ee where $r_{\rm s}$ is the black hole's Schwarzschild radius \cite{Landau}. 
Indeed, this is the case in classical general relativity, but one might ask whether it will still be valid for a quantum field. It is known that the singularity requires quantum gravitational treatment, however, as we are here taking the singularity seriously, the spacetime inside the black hole is in fact singular. Now we recall that quantum field theory is relativistic, thus no propagation outside the light cone is allowed. Moreover, we know that after entering the black hole, the trajectories avoiding the singularity correspond to superluminal propagation \cite{Landau}, therefore the fate of any field inside the black hole is to end up at the singularity inescapably. For a more detailed field theoretic (holographic) treatment of this issue we refer the reader to \cite{Thorac2}. There the time to reach the singularity slightly differs from the classical geodesics, but qualitatively nothing changes.\\

Let $\ket{\Psi}$ be the Hawking state of the whole radiation system \cite{Hawking, Fabbri}, namely the pairs created at the black hole horizon, 
\be
\hspace{-0.4cm}
 \ket{\Psi} = \bigotimes_\omega c_\omega \sum_N {\rm e}^{-\frac{ N \pi \omega} { \kappa} } \ket{ N_\omega} ^{\rm out} \otimes \ket{ N_\omega}^{\rm int}, 
 \ee
 where $c_\omega \equiv \sqrt{1- {\rm e}^{-8 \pi \omega M} }$ is a normalization factor, $N_\omega$ is the number of particles of energy $\omega$, while ``int'' and ``out'' label the Hilbert spaces for the particles falling inside the black hole and the particles escaping to the future infinity, respectively. 

 \be
 \hspace{-0.4cm}
\ket{\Psi}_t = \bigotimes_\omega c_\omega \sum_N {\rm e}^{- \frac{ N \pi \omega} { \kappa} } \ket{ N_\omega} ^{\rm out}_t \otimes \ket{ N_\omega}_t^{\rm int}, 
\ee
Once an ``int'' state with negative energy is created, it ends up at the singularity where it gets annihilated.

\subsection{Pure states inside and outside} 
We shall first consider the scenario that directly renders the outgoing radiation pure but unfortunately allows for non-local signaling.
The state after the particle hits the singularity is:
{\small
\be
&& \hspace{-0.3cm}  {\ket \Psi^\prime}_{t_s} = S^{\rm int} {\ket\Psi}_{t_s} \label{PsiA} 
\nonumber 
\\&& \hspace{-0.3cm}  
\equiv \! \ket{ r=0} \bra{r=0} \! \left( \! \bigotimes_\omega c_\omega \sum_N {\rm e}^{- \frac{ N \pi \omega} { \kappa} }\ket{ N_\omega }^{\rm out} _{t_s} \! \right) 
 \otimes \ket{ N_\omega}^{\rm int}_{t_s} 
  \nonumber  
 \\
&& \hspace{-0.3cm}   =  \bigotimes_\omega c_\omega \sum_N {\rm e}^{- \frac{ N \pi \omega} { \kappa} }\ket{ N_\omega }^{\rm out}_{t_s} \! \otimes \left( \! \ket{ r=0}  \bra{r=0}\ket{ N_\omega }^{\rm int} _{t_s} \right)  , 
\ee
}
\hspace{-0.25cm}
where $\bra{r=0} \equiv\bra{r=0}^{\rm int}$. 
After the time $t_s$ is passed we are sure with probability one that 
all the $\ket{N_\omega}^{\rm int}$ particles have evolved in to the point $r=0$ and they are crushed at the singularity. Indeed, by the definition of spacetime singularity, which we are here taking very seriously from the physical point of view, there is no dynamics beyond $r=0$ and any ``int'' particle must stop there without having any chance to escape because of the causal structure of the black hole interior. Therefore, 
\be
\boxed{  \bra{r=0}\ket{ N_\omega }^{\rm int}_{ t_s} = {\rm e}^{i \theta} } \, , \,\, \forall \, N \,\,\, {\rm and} \,\,\,\, \forall \, \omega ,
 \label{important}
 \ee
and the state in Eq.~(\ref{PsiA}) turns into 
\be
{\ket \Psi}^\prime = \left( \bigotimes_\omega c_\omega \sum_N  {\rm e}^{i \theta}  {\rm e}^{- \frac{ N \pi \omega} { \kappa} }\ket{ N_\omega }^{\rm out} \right)_{t>t_s}\! \! \! \! \! \! \! \! \!  \otimes \, \ket{r=0}^{\rm int} \, . 
\label{finalTensor}
\ee
 Note that the label ${t_s}$ is to indicate that we are not saying that all $ \ket{N_\omega^{\rm int}}$ 
 states are parallel to $\ket{r=0}$, this is impossible because for different $N$ they are orthogonal to each other. Instead, we are saying that all $ \ket{N_\omega^{\rm int}}$ 
 states evolve to the state $\ket{r=0}$ after an amount of time $t_s$ has passed. 
It is clear that the entangled state ${\ket \Psi}$ has evolved {\em deterministically} to a product state ${\ket \Psi}^\prime$ because 
the sum over $N$ does not involve ${\ket{N_\omega}}^{\rm int}$ anymore as a consequence of Eq.~(\ref{important}). Therefore, the entanglement is broken and there is no monogamy problem as well as no drama at the event horizon while the state outside is clearly pure.
Since the ``int'' state evolves to $\ket{r=0}$ regardless of what $\ket{N^{\rm int}}$ is as evident in Eq.~(\ref{finalTensor}). Therefore, there is no dependence on $N$ in the final state of the infalling Hawking particles.

Notice that the evolution to the singularity point is not unitary because the map (\ref{important}) to the singularity state is a ``many to one correspondence''. The black hole singularity seems to act as an observer making a measurement with a deterministic outcome. However, this non-unitary evolution at the singularity is supposed to have no observable consequences outside the black hole as argued in \cite{Thorac2, Iran}. Another issue is the quantum causality violation due to the singularity that we will discuss in the next section.

\subsubsection*{Causality violation}
It is well known that quantum mechanics is nonlocal, but it is impossible to send information faster than light ensuring that we never have a causality violation. 
In this section we show that in presence of a singularity, naked or not, in the way treated above, quantum mechanics violates causality. 
Here is our ``thought experiment''. 

Let us say Alice and Bob 
decide to explore the interior of a black hole to check for example whether it is dangerous or not. They prepare a bunch of EPR states, say 
\be
\ket{\psi} = \frac{1}{\sqrt{2}} \left( \ket{0_A 1_B} + \ket{1_A 0_B} \right) .
\ee
 Alice, the brave girl, jumps inside with one qubit of each EPR pair, while Bob stands outside waiting and holding the other bit of each pair. Having a spaceship, Alice can choose to arrive earlier  to the singularity. Before Alice (and her qubits) reach the singularity, the outside particles with Bob are in a maximally mixed state, namely  
 \be
 \rho_B = \frac{1}{2} \ket{0} \bra{0}_B + \frac{1}{2} \ket{1}\bra{1}_B.
 \label{safe}
 \ee
 Each reduced density matrix is proportional to the identity and 
 the outcome of Bob's measurement is highly random. On the other hand, after Alice throws the qubits into the singularity, the state with Bob, as we showed in the previous section, is pure,
\be
\rho_B = \frac{1}{2} \left( \ket{1} + \ket{0} \right)\left( \bra{1} + \bra{0} \right)_B , 
\label{danger}
\ee
and its statistics are different than the maximally mixed state. Therefore, Alice can inform Bob about the black hole interior by speeding up and reaching the singularity earlier with respect to $t_{\rm s}$. Indeed, they initially agreed on which state corresponds to a safe black hole and which state to a dangerous one: if Bob after an amount of time $t_{\rm s}$ observes the state (\ref{safe}) the black hole is a fine place, if he observes (\ref{danger}) then the black hole is dangerous.

\subsubsection*{Proper time shorter than the thermalization time} 
In the context of checking whether a state that falls into a black hole can be recovered from the radiation, in \cite{Susskind, PP, Thorac2} the authors argued that, in case it is possible, it takes at least a thermalization time of the order 
\be
{\rm max} \, \left\{ r_{\rm s} \ln \frac{r_{\rm s}}{l_P}, \,\,  r_{\rm s}  \right\}. 
\ee
This time is longer than the proper time to reach the singularity (\ref{timesing}) 
 ($l_P$ is the Planck length). Therefore, the ``int" and ``out" pairs will be disentangled and the ``out'' particles will have the chance to be entangled outside the black hole in a Page-like scheme before the thermalization time.

Finally, we can summarize the content of this subsection as follows. The ``int'' particles are all reset to the singularity state  in a very short proper time. Therefore, the Hawking state evolves fast and non-unitarily to a pure state. This scenario violates quantum causality because we can in principle communicate with the black hole exterior using a bunch of EPR pairs.

\subsection{Breaking the entanglement without causality violation}
An alternative model avoids causality violation. Looking at the fact that everything ends up stuck at the singularity point, this effect can be represented as an operator $\boldsymbol{\varepsilon }$ that maps any state crossing the event horizon onto one point (the singularity itself) after a finite amount of time $t_{\rm s}$.

Whatever initial state, represented by a reduced density matrix $\rho_{\rm int}$, should be mapped to the singularity at $r=0$. We shall later consider smearing around this position, but for now we treat this strictly as 
\be
\rho_{\rm int} \,\, \rightarrow \,\, \ket{r=0}\bra{r=0} \, 
\ee
for all allowed $\rho_{\rm int}$. Since it is deterministic it happens with probability one. This fully defines the quantum map representing the evolution, namely 
\be
\boldsymbol{\varepsilon }(\rho_{\rm int}) = \ket{r=0} \bra{r=0}, \: \: \forall \, \rho_{\rm int} \, ,
\ee
We shall refer to the above as the singularity map, which as a many-to-one map is not unitary.
Now, facing such an unusual map, one would like to check whether it is standard in the quantum information theoretic sense, that is, whether there is a completely positive trace-preserving (CPTP) map that can do the same job. Indeed one can easily check that this map is in fact equivalent to unitarily swapping the input system with a system in the state $\ket{r=0} \bra{r=0}$, by applying the unitary $U_{\rm swap}$ such that $$U_{\rm swap}\ket{i}_A\ket{j}_B=\ket{j}_A\ket{i}_B\,\,\forall B$$
and then tracing over the second system $B$. 
Equivalently $U_{\rm swap}$ moves the trace from the $\ket{r=0} \bra{r=0}$ system to the ``int'' system, 
\be
\boldsymbol{\varepsilon }(\rho_{\rm int}) &=& \Tr_{\rm sing} \left\{ U_{\rm swap} \big( \rho_{\rm int} \otimes \ket{r=0}\bra{r=0} \big) U^+ _{\rm swap} \right\} \nonumber \\ 
&=& \Tr_{\rm int} \{\ket{r=0}\bra{r=0} \otimes \rho_{\rm int} \} \nonumber \\
&=& \ket{r=0}\bra{r=0} \, . 
\ee
The existence of the above description, in terms of a unitary interaction with another system followed by tracing out, shows that this is a completely positive (CP) map, in the language of quantum maps~\cite{Nielsen}. Moreover, it is trace preserving (TP) by inspection ($\Tr \ket{r=0}\bra{r=0}=\Tr \rho_{\rm int}=1$). Thus it is a CPTP map, a standard set of maps. The data processing inequality \cite{Nielsen} applies to CPTP maps, implying here that the singularity map cannot increase the mutual information (nor entanglement) with the outside. 
More generally one sees that $\boldsymbol{\varepsilon }(\rho_{\rm int})$, which is the state of anything crossing the event horizon, can not be correlated with any other state because inside it is mapped to a pure state.

To summarize, in this subsection we constructed a non-unitary CPTP map that resets the density matrix for the ``int'' state onto the singularity, breaking the entanglement between the ``int'' and ``out'' states, but without violating causality.

\subsubsection*{Pure states cannot be correlated with others}
Let us consider the following systems. 
The system $A$ in a pure state, an arbitrary system 
$B$ and the whole system $AB \equiv  A \cup B$ that does not need to be in a pure state.

For completeness, we are now going to show that the mutual information of $A$ and $B$, defined by
\be
S(A:B)= S(A) + S(B) - S(AB),
\ee
is identically zero.

Since $A$ is a pure state then the entropy $S(A)=0$. 
Now, applying the sub-additivity rule to the systems $A$ and $B$ we find 
\be
S(AB) \leq S(A) + S(B) \leq S(B) \, .
\label{S2}
\ee
Let us consider an additional system, $R$, which purifies $AB$ (namely the whole system $ABR$ is pure.)  Therefore, $S(ABR) =0$ or equivalently it is known that 
\be
S(AR) = S(B) \quad  {\rm and} \quad S(R)= S(AB) \, .
\label{S3}
\ee
Applying again the sub-additivity to the system $AR$ we get 
\be
S(AR) \leq S(R) + S(A)  \, . 
\label{S4}
\ee
Substituting (\ref{S3}) in (\ref{S4}) we obtain 
\be
S(AB) \geq S(B) - S(A) = S(B) 
\label{S5}
\ee
because $A$ is pure. 
Combining (\ref{S2}) and (\ref{S5}) we end up with 
\be
S(B)= S(AB) 
\ee
that finally implies 
\be
S(A:B)= S(B) - S(AB) = 0. 
\ee
So, 
there is no mutual information between any system $A$ in a pure state and any other system $B$. Therefore, the ``int'' radiation, after being set pure by the singularity, cannot be entangled with the outside radiation anymore.
One can go through the above argument for a finite $S(A)$ to find more generally that 
\be
\label{eq:mutinfbound}
S(A:B)\leq 2 S(A),
\ee
so the argument generalizes in that sense. 
This then means that as soon as the infalling states reach the singularity their entanglement with the outside is finally broken, thus allowing for the possibility of outgoing radiation particles being correlated with each other without violating monogamy.\\

\section{The case of spread singularities}
This approach to the information paradox treats the singularity as a single point with an exactly known position. This assumption might not be very plausible in quantum mechanics. 
In this paragraph we generalize the firepoint to a singular extended region with associated  probabilistic distribution. This means the singularity is still a singularity, but it exists in different positions with different probability amplitudes.
We should clarify that we are not considering here any resolution of the singularity problem. 
The singularity is still an essential one, but it is probabilistically distributed on a discrete set of points or on an extended manifold. For example, the singularity could be in the following state in the position space
\be
\ket{0} = \sum_{i} \lambda_i \ket{r_{i}} \, .
\ee
The generalization to the continuum case is trivial.

{\small
\be
&& \hspace{-0.4cm}  {\ket \Psi^\prime}_{t_s} = S^{\rm int} {\ket\Psi}_{t_s} \label{PsiA7} 
\nonumber 
\\&& \hspace{-0.4cm}  
= \! \!  \sum_{i} \lambda_i \ket{r_{i}}\sum_{j} \lambda_j \bra{r_{j}} \! \left( \! \bigotimes_\omega c_\omega \sum_N {\rm e}^{- \frac{ N \pi \omega} { \kappa} }\ket{ N_\omega }_{t_s}^{\rm out} \! \right) 
 \otimes \ket{ N_\omega}_{t_s}^{\rm int} 
  \nonumber  
 \\
&& \hspace{-0.4cm}   =\!    \bigotimes_\omega c_\omega \! \! \! \sum_N {\rm e}^{ \frac{ - N \pi \omega} { \kappa} }\ket{ N_\omega }^{\rm out} \! \! \otimes \! \! \left( \! \! \sum_{i} \! \lambda_i \ket{r_{i}}\! \! \sum_{j} \lambda_j \bra{r_j}\ket{ N_\omega }^{\rm int}_{ t_s} \right) \! .
\ee
}
\hspace{-0.3cm}
Now 
$\sum_{j} \lambda_j \bra{r_j}\ket{ N_\omega }^{\rm int}_{ t_s}$
is exactly the probability amplitude of finding the state $\ket{ N_\omega }^{\rm int} _{t_s}$ at the singularity after $t_s$ passes.
This probability amplitude as argued before is nothing but a phase $e^{i \theta}$.
Then 
\be
\hspace{-0.2cm}
{\ket \Psi}^\prime = \left( \bigotimes_\omega c_\omega \sum_N {\rm e} ^{i\theta} {\rm e}^{- \frac{ N \pi \omega} { \kappa} }\ket{ N_\omega }^{\rm out} \right) \! \otimes  \! \sum_{i} \lambda_i \ket{r_{i}}^{\rm int} \!
\! . 
\label{finalTensor2}
\ee
Again, we clearly ended up in a pure state for the black hole exterior, and signaling is still possible.\\

Now the singularity map acts in the following way, 
\be
&&\hspace{-0.5cm}
 \boldsymbol{\varepsilon }(\rho) 
 = \Tr_{\rm sing} \Big\{ U_{\rm swap} \nonumber \\
 &&\hspace{-0.5cm}
  \Big( \sum_{N,N^\prime} c_{N, N^\prime} \underbrace{ \ket{N_{\rm int}, 
N_{\rm out} } \bra{ N{^\prime}_{\rm int}, N^{\prime}_{\rm out}} }_{\rho(N, N^\prime)}  
 \otimes \ket{0} \bra{0} \Big) U^+ _{\rm swap} \Big\}  \nonumber \\
&& \hspace{-0.5cm}
= \Tr_{\rm int} \Big\{  \ket{0} \bra{0} \otimes 
  \sum_{N,N^\prime} c_{N, N^\prime} \, \rho(N, N^\prime)    \Big\}  \nonumber \\
&&\hspace{-0.5cm}
= \ket{0} \bra{0} \otimes \sum_N \rho_{\rm out} (N) 
\nonumber \\
&& \hspace{-0.5cm}
 = \sum_{i,j} \lambda_i \bar{\lambda}_j \ket{r_{i}} \bra{ r_{j}} \otimes \underbrace{\sum_N \rho_{\rm out}(N)}_{\rho_{\rm out}} \, , 
 \ee
where $U_{\rm swap}$ again swaps the ``int'' state and the ``singularity'' state, while $\rho_{out}$ is the reduced density matrix of the ``out'' radiation. This is a standard map in quantum theory, the inside is traced out and the outside might still be in a mixed state, but what is new is that the inside is mapped to a pure state, so now if the outside particles get entangled with each other they do not violate monogamy

\section*{Conclusions}
We affronted the information loss problem from a very conservative point of view without completely abandoning the firewall idea, which has been here identified with the spacetime singularity in $r=0$ in the exact Schwarzschild solution of Einstein's gravity.   
The singularity can take active part in the information loss problem in two similar, but slightly different ways. 
The first one reads as follows. 
The singularity can play the role of a firepoint (firewall of zero radius) capable of manifestly purifying the states outside the horizon by resetting the inner states that feel the ``position'' of the singularity.
However, this ``purification at a distance'' allows for communication with the world outside the black hole, violating causality. 
We could in principle send a space probe equipped with a bunch of EPR pairs inside the black hole and get information about the interior of the black hole from the spacetime region outside the event horizon. 

Alternatively, we can model the impact of the singularity in the Hawking's evaporation process through a CPTP map that  
breaks the entanglement between ingoing and outgoing particles thus opening the possibility of refuting the monogamy issue that has disproved the Page's argument. In this case we do not have causality violation because the singularity itself maps the ``int'' density matrix to $r=0$ yet keeping the density matrix of the ``out'' radiation intact.

Since in both cases entanglement can be broken avoiding violations of the monogamy theorem, we think that the Page's argument might still be a viable solution of the information loss problem.

\section*{Appendix}

We review the radial geodesic motion of a massive test-particle. On the base of the geodesic principle the action reads:
\be
&& S_{\rm cp} = -m \int ds = - m \int \sqrt{ - {g}_{\mu\nu} d x^\mu d x^\nu} \nonumber \\
&& \hspace{0.65cm} =  - m  \int \sqrt{ - g_{\mu\nu} \frac{d x^\mu}{d \lambda} \frac{d x^\nu}{d \lambda} } 
\, d \lambda \, , 
\label{Spc}
\ee
where $\lambda$ is an affine parameter, and $x^\mu(\lambda)$ is the trajectory of the particle. Therefore, the Lagrangian is:
\be
L_{\rm cp}  =  - m \sqrt{ -  g_{\mu\nu} \dot{x}^\mu \dot{x}^\nu } \, , 
\ee
and the translation invariance in the time-like coordinate $t$ implies 
\be
\frac{\partial L_{\rm cp}}{\partial \dot{t} } = - \frac{m^2  g_{tt} \dot{t}}{L_{\rm cp}} = {\rm const.} = - E \, \Rightarrow \,   \dot{t} = \frac{L_{\rm cp} E }{m^2 g_{tt}} . 
\label{ConstE}
\ee
Since we are interested in evaluating the proper time of the particle necessary to reach the point $r=0$, we must choose the proper time gauge, namely $\lambda = \tau$. In this case, $E$ is the energy of the test-particle and
\be
\frac{d s^2}{d \tau^2} = - 1 
\quad \Longrightarrow \quad  L_{\rm cp}= - m
\,\, \Rightarrow \,\,  \dot{t} = - \frac{ E }{m \, g_{tt}} \, .
\label{PTG}
\ee
Replacing~(\ref{ConstE}) in $g_{\mu\nu} \dot{x}^\mu \dot{x}^\nu = - 1$, we end up with 
\be
&& g_{tt} \dot{t}^2 + g_{rr} \dot{r}^2  = -1 \nonumber \\
&& \dot{r}^2 = - \left( 1- \frac{2 M}{r}  \right)  + \frac{E^2}{m^2} \, . 
\ee
For a particle at rest at infinity $E =m$ and the above equation simplifies to 
\be
\dot{r}^2 = \frac{2M}{r} \, .
\label{Es}
\ee
For a particle traveling towards the singularity starting from the radial coordinate $r_0$ for $\tau =0$, 
\be
\frac{dr}{d \tau} = - \sqrt{\frac{2 M}{r}} \,\,\, \Longrightarrow \,\,\, \tau = \frac{2}{3} \left( \frac{r_0^{\frac{3}{2}}}{r_{\rm s}^{\frac{1}{2}}} -  \frac{r^{\frac{3}{2}}}{r_{\rm s}^{\frac{1}{2}}} 
\right) \, , 
\ee
where is the Schwarzschild radius $r_{\rm s} = 2 M$. 
Therefore, the proper time to reach the singularity ($r=0$) starting from the event horizon ($r=r_{\rm s}$) is:
\be
\tau_s = \frac{2 r_{\rm s}}{3} =  \frac{4 M G}{3 c^3} \, .
\ee

It is well known that any particle inside the event horizon is forced to reach the singularity in $r=0$ because of the causal structure of the spacetime. 
Therefore, we conclude that any massive particle (we can prove the same for massless particles in the affine parametrization of the geodesics) reach the singularity in very short time.


\begin{thebibliography}{99}
\bibitem{Hawking}
  S.~W.~Hawking,
  ``Particle Creation by Black Holes,''
  Commun.\ Math.\ Phys.\  {\bf 43}, 199 (1975)
  Erratum: [Commun.\ Math.\ Phys.\  {\bf 46}, 206 (1976)].
\bibitem{Fabbri} Alessandro Fabbri, Jose Navarro-Salas, Modeling Black Hole Evaporation, Imperial College Press, 2005.
\bibitem{Page}
  D.~N.~Page,
  ``Average entropy of a subsystem,''
  Phys.\ Rev.\ Lett.\  {\bf 71}, 1291 (1993)
  [gr-qc/9305007].
\\
  D.~N.~Page,
  ``Information in black hole radiation,''
  Phys.\ Rev.\ Lett.\  {\bf 71}, 3743 (1993)
  [hep-th/9306083].
 
\bibitem{Mathur} 
  S.~D.~Mathur,
  ``The Information paradox: A Pedagogical introduction,''
  Class.\ Quant.\ Grav.\  {\bf 26}, 224001 (2009)
  [arXiv:0909.1038 [hep-th]].


\bibitem{coffman} 
  V.~Coffman, J.~Kundu and W.~K.~Wootters,
  ``Distributed entanglement,''
  Phys.\ Rev.\ A {\bf 61}, 052306 (2000)
  [quant-ph/9907047].


\bibitem{Braunstein} 
  S.~L.~Braunstein, S.~Pirandola and K.~Życzkowski,
  ``Better Late than Never: Information Retrieval from Black Holes,''
  Phys.\ Rev.\ Lett.\  {\bf 110}, no. 10, 101301 (2013)
  [arXiv:0907.1190 [quant-ph]].



\bibitem{AMPS} 
  A.~Almheiri, D.~Marolf, J.~Polchinski and J.~Sully,
  ``Black Holes: Complementarity or Firewalls?,''
  JHEP {\bf 1302}, 062 (2013)
  [arXiv:1207.3123 [hep-th]].
  
\bibitem{Thorac1} 
  D.~A.~Lowe and L.~Thorlacius,
  ``Black hole complementarity: The inside view,''
  Phys.\ Lett.\ B {\bf 737}, 320 (2014)
  [arXiv:1402.4545 [hep-th]].




\bibitem{Thorac2}
  D.~A.~Lowe and L.~Thorlacius,
  ``Quantum information erasure inside black holes,''
  JHEP {\bf 1512}, 096 (2015)
  [arXiv:1508.06572 [hep-th]].
\bibitem{Complementarity}
  L.~Susskind, L.~Thorlacius and J.~Uglum,
 ``The Stretched horizon and black hole complementarity,''
  Phys.\ Rev.\ D {\bf 48}, 3743 (1993)
  [hep-th/9306069].
\bibitem{Iran} 
  F.~S.~Dündar and K.~Hajian,
  ``Quantum Jump from Singularity to Outside of Black Hole,''
  JHEP {\bf 1602}, 175 (2016)
  [arXiv:1511.03572 [gr-qc]].
\bibitem{Landau} L.D. Landau and E.M. Lifshitz, The Classical Theory of Fields, Elsevier Ltd 2007.
\bibitem{Bekenstein} 
  J.~D.~Bekenstein,
  ``How does the entropy / information bound work?,''
  Found.\ Phys.\  {\bf 35}, 1805 (2005)
  [quant-ph/0404042].
\bibitem{Nielsen} Nielsen and Chuang, Quantum Computation and Quantum Information, Cambridge, 2000.
\bibitem{Susskind}
  L.~Susskind and L.~Thorlacius,
  ``Gedanken experiments involving black holes,''
  Phys.\ Rev.\ D {\bf 49}, 966 (1994)
  [hep-th/9308100].
\bibitem{PP}
  P.~Hayden and J.~Preskill,
  ``Black holes as mirrors: Quantum information in random subsystems,''
  JHEP {\bf 0709}, 120 (2007)
  [arXiv:0708.4025 [hep-th]].
\end{thebibliography}
\end{document}